\documentclass[letterpaper]{article} 
\usepackage{aaai2026}  
\usepackage{times}  
\usepackage{helvet}  
\usepackage{courier}  
\usepackage[hyphens]{url}  
\usepackage{graphicx} 
\usepackage{natbib}  
\usepackage{caption} 
\usepackage{algorithm}
\usepackage{algorithmic}

\usepackage[utf8]{inputenc} 
\usepackage[T1]{fontenc} 
\usepackage{type1cm} 
\usepackage{amsmath} 
\usepackage{booktabs} 
\usepackage{bold-extra} 
\usepackage{graphicx} 
\usepackage{hyphenat} 
\usepackage{mathtools} 
\usepackage{microtype} 
\usepackage{multirow} 
\usepackage{subcaption} 
\usepackage{siunitx} 
\usepackage{xfrac} 
\usepackage{xspace} 
\usepackage{cleveref} 

\newcommand{\spara}[1]{\smallskip\noindent\textbf{#1}}

\newenvironment{squishlist}
{\begin{list}{$\bullet$}
{\setlength{\itemsep}{0pt}
\setlength{\parsep}{3pt}
\setlength{\topsep}{3pt}
\setlength{\partopsep}{0pt}
\setlength{\leftmargin}{1.5em}
\setlength{\labelwidth}{1em}
\setlength{\labelsep}{0.5em}}}
{\end{list}}

\usepackage[bb=dsserif]{mathalpha} 
\usepackage{bm} 
\usepackage{newfloat}
\usepackage{listings}
\DeclareCaptionStyle{ruled}{labelfont=normalfont,labelsep=colon,strut=off} 
\floatstyle{ruled}
\newfloat{listing}{tb}{lst}{}
\floatname{listing}{Listing}
\usepackage{xcolor}
\newcommand{\answerYes}[1]{\textcolor{blue}{#1}} 
 
\newcommand{\answerNA}[1]{\textcolor{gray}{#1}}

\title{Effects of Mainstream Visibility on Conspiracy Communities:\\Reddit after Epstein's `Suicide'}
\author{
    Asja Attanasio\textsuperscript{\rm 1},
    Francesco Corso\textsuperscript{\rm 1, \rm2},
    Gianmarco De Francisci Morales\textsuperscript{ \rm2, \rm3},
    Francesco Pierri\textsuperscript{\rm 1}
}
\affiliations{
    \textsuperscript{\rm 1}DEIB, Politecnico di Milano, Via Golgi 34, Milano, Italy\\
    \textsuperscript{\rm 2}CENTAI, Corso Inghilterra 4, Torino, Italy\\
    \textsuperscript{\rm 3}ISPIC, Corso Inghilterra 4, Torino, Italy\\
    asja.attanasio@mail.polimi.it, francesco.corso@polimi.it,
    gdfm@acm.org,
    francesco.pierri@polimi.it,\\
    \noindent\textcolor{red}{\textbf{Please check the published version of the paper at AAAI ICWSM 2026}}

}

\usepackage{bibentry}

\begin{document}

\maketitle

\begin{abstract}
Following the death of Jeffrey Epstein, the subreddit r/conspiracy experienced a significant visibility shock that brought mainstream users into direct contact with established conspiracy narratives.
In this work, we explore how large-scale surges in public attention reshape participation and discourse within online conspiracy communities.
We ask whether a sudden increase in exposure changes who join r/conspiracy, how long they stay, and how they adapt linguistically, compared with users who arrive through organic discovery. 
Using a computational framework that combines toxicity scores, survival analysis, and lexical and semantic measures over a period of 12 months, 
we observe that mainstream visibility is is associated with patterns consistent with a selection mechanism rather than a simple amplifier.
Users who join the conspiracy community during the arrest-period tend to show higher linguistic similarity to core users, especially regarding linguistic and thematic norms and showing more stable engagement over time. 
By contrast, users who arrive during the height of public visibility remain semantically distant from core discourse and participate more briefly. 
Overall, we find that mainstream visibility is connected with changes in audience size, community composition, and linguistic cohesion.
However, incidental exposure during attention shocks does not typically produce durable, integrated community members.
These results provide a more nuanced understanding of how external events and platform visibility influence the growth and evolution of conspiracy spaces, offering insights for the design of responsible and transparent recommendation systems.
\end{abstract}

\section{Introduction}
Conspiracy theories are false or unverified narratives that allege secret,
often sinister plots or events involving groups, organizations, or governments \cite{douglas2023conspiracy}.
They pose significant risks since they can change individuals' perception of real events \cite{uscinski2017climate}.
Social media platforms provide especially fertile ground for their circulation, in part due to the emergence of fringe online communities \cite{rollo2022communities,spohr2017fake}.
Such theories range from relatively benign conjectures to more harmful beliefs that amplify misinformation, heighten polarization, and potentially lead to serious real-world consequences.

Online conspiracy communities are critical environments for understanding how digital platforms shape ideological discourse.
Yet, it's not clear how mainstream visibility of these fringe communities plays a role in determining who joins these communities, how long they stay, and whether they culturally integrate~\cite{klein_pathways_2019, phadke_pathways_2022, venturini_online_2022, zeng_conspiracy_2022,corso2025early}. 
Existing work on conspiracy engagement shows that users follow distinct pathways into and through conspiracy spaces and that linguistic adaptation and community ties play a central role in deeper involvement~\citep{klein_pathways_2019,phadke_pathways_2022}.
At the same time, research on platform interventions and deplatforming shows that conspiracy communities can be unusually resilient and mobile, rebuilding networks elsewhere even after bans~\citep{monti_online_2023}.
What remains unclear is how platform-level visibility itself, before any formal intervention, shapes who enters these communities and what kinds of membership emerge.

Reddit offers a useful setting for such a research question~\cite{proferes_studying_2021}.
It organizes content into topical \emph{subreddits} and uses ranking and recommendation systems to decide which posts appear on the site's homepage. 
When a subreddit reaches the homepage, it exposes its community to millions of users who are not subscribed to it and might not otherwise encounter its content. 
The subreddit \texttt{r/conspiracy} is Reddit's largest conspiracy community, with more than two million members and a long record of discussions that range from historical events to contemporary politics. 
This combination of scale, public visibility, and stable infrastructure makes \texttt{r/conspiracy} an ideal case for studying how visibility events affect community composition, participation, and language~\citep{de_wildt_participatory_2024}.

The death of financier Jeffrey Epstein on August 10, 2019, created a sharp visibility shock~\citep{schatto-eckrodt_seed_2024}. 
The peculiar circumstances of his death aligned fringe conspiracy discourse with mainstream skepticism and brought \texttt{r/conspiracy} and its 2 million members to the fore, reaching the platform's homepage, as reported by users from the same community.\footnote{\url{https://reddit.com/r/conspiracy/comments/cojrq9}}\footnote{\url{https://reddit.com/r/conspiracy/comments/covvj0}}
This visibility event changed the community's reach without changing its formal rules or moderation structures, and it did so over a short, well-defined time window as visible in \Cref{fig:app_users_active}

In this work, we examine how Reddit's homepage visibility after Epstein's death is associated with changes in community composition consistent with a compositional filter that reshapes \texttt{r/conspiracy}'s user base, participation patterns, and linguistic integration. 
We focus on whether users who arrive during this visibility event behave and write differently from those who discover the community through more organic pathways, and whether these differences persist over time. 
By combining behavioral, lexical, and semantic measures, this study links discovery context to toxicity, retention, and linguistic alignment.
Specifically, we aim to answer the following research questions:

\begin{squishlist}
\item[\textbf{RQ1:}] How do sudden visibility events alter the behavioral and linguistic signatures of established community members?
\item[\textbf{RQ2:}] How does homepage visibility affect community composition and user retention patterns?
\item[\textbf{RQ3:}]Does the timing and context of community entry influence linguistic integration patterns?
\end{squishlist}

For \textbf{RQ1}, we use an Interrupted Time Series (ITS) analysis of toxicity levels and an Empath lexical category analysis of thematic shifts. 
Our analysis reveals that the core users' language changed significantly at the moment of Epstein's death, and their toxicity decreased significantly.
Moreover, we measure a shift in the topics discussed, as mentions of \textit{violence} and \textit{crime} increased significantly, while \textit{government} and \textit{religion} are reduced.  

For \textbf{RQ2}, we construct four mutually exclusive user cohorts based on temporal boundaries and engagement focus, and we use Kaplan-Meier survival analysis to model retention trajectories and compare users who enter during the heightened visibility of the death-period versus those who arrived during the earlier arrest-period.
Users whose first post is about Epstein show higher retention compared to users whose first engagement is not related to Epstein, thus indicating that the Epstein suicide topic possibly attracted individuals with a stronger interest in conspiracy theories.

Finally, for \textbf{RQ3}, we employ Sentence-BERT embeddings to compute semantic similarity between newcomer posts and established community discourse, and we track whether users who arrive via homepage visibility exhibit persistent linguistic divergence from community norms. 
Users who join during the arrest have significantly higher linguistic similarity to the core users of the community. 
This signal remains strong across all the observation windows, while the similarity of users joining after Epstein's death is lower across the study period.
These findings suggest that users who discover the conspiracy community during the arrest period show a stronger language adaptation to long-term conspiracy theorists, potentially signaling a stronger intrinsic interest in the community.

Taken together, these findings connect a concrete visibility event to broader debates about algorithmic exposure, community growth, and online radicalization, and they provide evidence about when visibility changes who appears in a community versus when it changes how that community evolves.

\section{Related Work}
Online conspiracy theories have been examined from multiple perspectives, spanning information diffusion, user trajectories, linguistic adaptation, and platform governance.

\spara{Diffusion of conspiracy theories on social media platforms.}
Research shows that social media provides fertile ground for conspiratorial content to circulate at scale~\cite{bessi2015science}.
During moments of crisis---such as the COVID-19 pandemic---these dynamics intensify, and conspiracy narratives often spread faster than verified information~\cite{kauk2021understanding}.
Propagation has been modeled using epidemiological frameworks, revealing higher virality and persistence relative to factual news.
Other work highlights cross-platform reinforcement, demonstrating that discourse emerging outside conspiracy spaces can seed and accelerate conspiracy narratives elsewhere~\cite{paudel2021soros}.
Together, these studies emphasize the need to understand how platform mechanisms facilitate the growth and reach of conspiracy content.

\spara{Detection of conspiracy theories online.}
Several studies have focused on detecting online conspiracy theories. 
combining natural-language processing, network analysis, and graph-based models. Recent studies showed that both supervised classifiers and large language models can detect conspiratorial messages with high accuracy, reaching high classification scores even in real settings \cite{pustet-etal-2024-detection,corso-etal-2025-conspiracy}. Recent work also showed how LLMs also can be used as proxy for simulating conspiracy-inclined individuals \cite{corso2025androids}. 
Another approach leverages ``event-relation graphs,'' mapping how events in a text are causally or temporally related; this helps flag articles whose narrative structure reveals irrational or conspiratorial linkages \cite{lei2023identifying}. 
Some frameworks further enrich detection by combining sentiment and topic modeling with qualitative coding, useful for uncovering subtle, implicit conspiratorial themes that keyword- or formula-based systems often miss~\cite{langguth2023coco}.

\spara{Conspiracy engagement and user pathways.}
Work on conspiracy engagement shows that users rarely arrive in conspiracy communities by accident.
\citet{klein_pathways_2019} trace Reddit users before their first post in \texttt{r/conspiracy} and show that future participants already display distinctive linguistic patterns and spend time in related subreddits. 
\citet{phadke_pathways_2022} follow users after they join \texttt{r/conspiracy} and identify multiple engagement trajectories, from stable high participation to rapid decline, linking persistent involvement to insider language and monological worldviews. 
Similarly, \citet{corso2025early} analyze the linguistic traits of users active in \texttt{r/conspiracy} in mainstream discourse, finding unique community-specific linguistic fingerprints that identify conspiracy users in the non-fringe communities, which hold strong predictive power even years before the user's first activity on \texttt{r/conspiracy}.
Together, these studies treat conspiracy engagement as a selective process shaped by prior dispositions and internal adaptation, but they focus on trajectories \emph{within} the community rather than on how mainstream visibility changes who arrives in the first place.

\spara{Linguistic adaptation and integration.}
Another line of research studies how users adapt their language to community norms. 
\citet{danescu-niculescu-mizil_no_2013} propose a lifecycle model in which users first align strongly with local norms and then enter a conservative phase where their style stabilizes and may drift from an evolving baseline. 
Studies of conspiracy discourse apply similar ideas to specific events and platforms. 
\citet{schatto-eckrodt_seed_2024} show how conspiracy narratives about Epstein's death build on older myths and political tensions, while work on \texttt{r/conspiracy} documents that dramatic events bring sudden influxes of new users and shift topics and tone~\citep{samory_conspiracies_2018, de_wildt_participatory_2024}. 
These studies demonstrate that external shocks matter for discourse, but they do not compare linguistic integration across cohorts defined by how users discover the community.

\spara{Platform interventions, visibility, and conspiracy communities.}
Several work examines how platforms intervene in problematic communities. 
Research on bans and deplatforming shows that removing communities does not always reduce their activity. 
\citet{monti_online_2023} find that QAnon users banned from Reddit migrate to alternative platforms and rebuild their networks, while \citet{trujillo_make_2022} show that quarantining \texttt{r/The\_Donald} briefly reduces activity and toxicity before both rebound. 
Other work documents spillover effects when users carry norms and content to fringe platforms after restrictions~\citep{horta_ribeiro_deplatforming_2023, russo_spillover_2023}. 
This literature treats visibility principally as something platforms reduce through sanctions and moderation tools.

Overall, these different works show that conspiracy engagement is selective, that users adapt linguistically to community norms, and that platform decisions can reshape participation. 
However, the role of sudden visibility surges in filtering community membership and reshaping discourse remains under-explored.
We address this by comparing newcomer cohorts arriving during two distinct attention spikes—the arrest-period and the high-visibility death-period—linking these arrival windows to core-user toxicity, newcomer retention, and linguistic integration to evaluate how large-scale interest shocks influence community dynamics.

\section{Methods}
We analyze all submissions and comments posted on \texttt{r/conspiracy} between January 1 and December 31 2019, collected from monthly Pushshift archives of Reddit~\citep{baumgartner_pushshift_2020}. 
After filtering out deleted authors and empty or deleted texts, the dataset contains \num{73320} submissions and \num{2207905} comments, for a total of \num{2281225} posts (\Cref{table:dataset-sizes}). 

\begin{table}[b]
    \label{table:dataset-sizes}
    \begin{center}
        \begin{tabular}{lS[table-number-alignment=center,table-format=7.0]}
        \toprule
        \textbf{Type} & \textbf{Number} \\
        \midrule
        Submissions & 73320 \\
        Comments & 2207905 \\
        \midrule
        Total posts & 2281225 \\
        \bottomrule
    \end{tabular}
    \end{center}
    \caption{Number of submissions, comments, and total posts after data cleaning.}
\end{table}

\begin{figure}[!t]
  \centering
  \includegraphics[width=\linewidth]{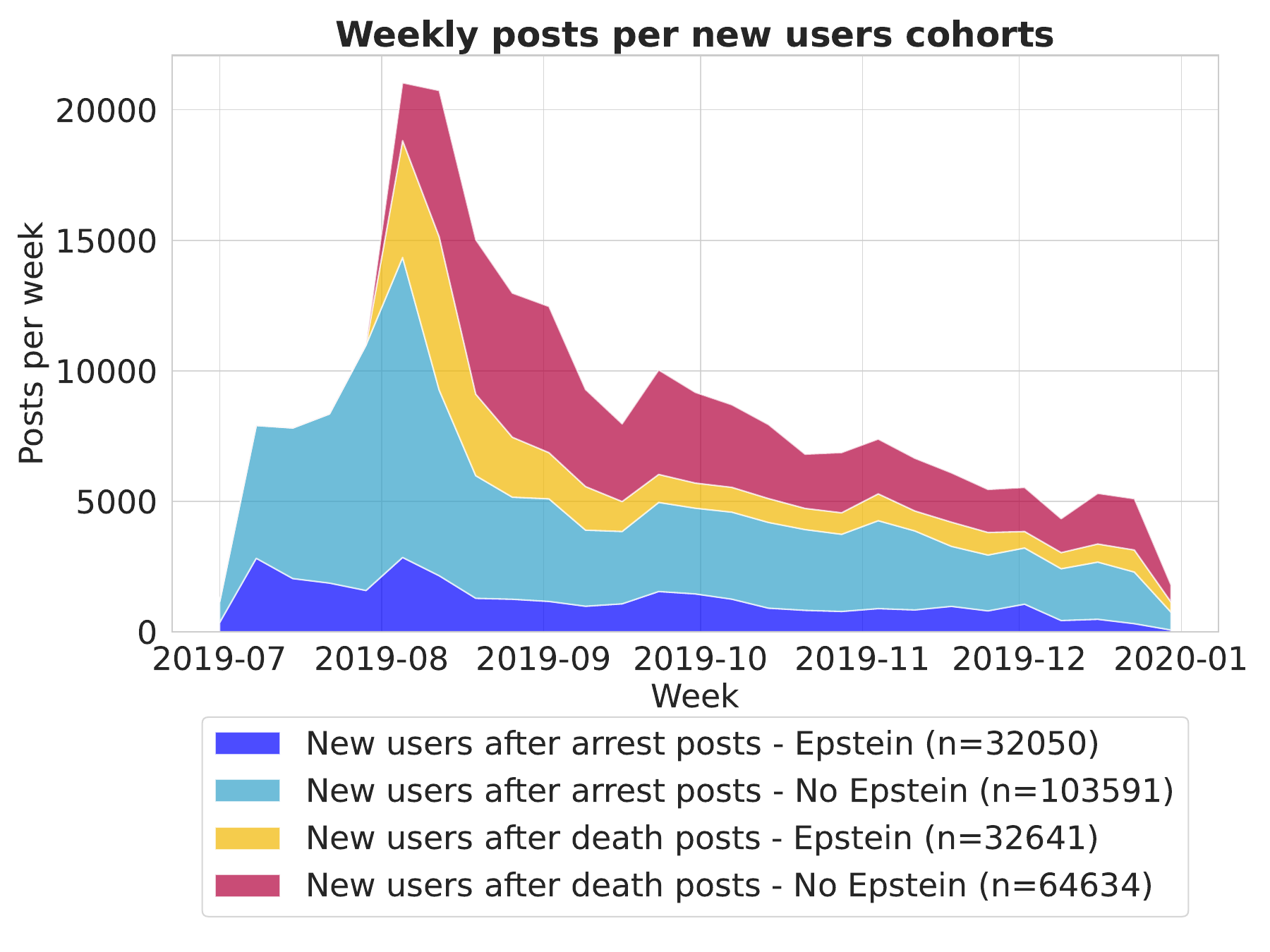}
  \caption{Longitudinal cohort analysis showing weekly posting activity for each of the four user cohorts.}
  \label{fig:cohorts_activity}
\end{figure}

We identify Epstein-related threads with a keyword-based matching approach over submission titles and texts on the following set of keywords: \texttt{epstein}, \texttt{jeffrey epstein}, \texttt{ghislaine}, or \texttt{maxwell}. 
We then include all comments to these submissions, which gives a focused set of posts that discuss Epstein either directly or inside those threads.
We chose the previously mentioned keywords due their strict connection with our case study, which helps us in maximizing the precision of the matching, even at the cost of a lower recall, to ensure the "engaged" cohorts are truly topical.
We define four mutually exclusive newcomer cohorts based on when users first post in \texttt{r/conspiracy} and whether that first post concerns Epstein. 
The cohorts are:

\begin{squishlist}
    \item[\emph{arrest--engaged}:] users whose first post appears between Epstein's arrest (6 July 2019) and his death (10 August 2019) in an Epstein-related thread ($n=\num{32050}$).
    \item[\emph{arrest--not-engaged}:] users whose first post appears between the arrest and the death, but not in an Epstein-related thread ($n=\num{103591}$).
    \item[\emph{death--engaged}:] users whose first post appears in the month after Epstein's death (11 August--10 September 2019) in an Epstein-related thread ($n=\num{32641}$).
    \item[\emph{death--not-engaged}:] users whose first post appears in the month after Epstein's death, but not in an Epstein-related thread ($n=\num{64634}$).
\end{squishlist}
\Cref{fig:cohorts_activity} shows the temporal activity of different cohorts.
We treat users who post in \texttt{r/conspiracy} before 6 July 2019 as core users and use them as the baseline for all toxicity, lexical, and semantic analyses.

Additionally, as shown in \Cref{fig:google_searches} in the Appendix, there are two distinct spikes in public interest (proxy via Google Search trends), and we use these events as anchors to examine corresponding activity on Reddit. 
Overall engagement on Reddit appears largely stable; therefore, the patterns we observe are likely driven by these attention spikes, which trigger shifts within the platform.

\subsection{Toxicity time series}

We study how the toxicity of core users changes around Epstein's death by constructing a daily toxicity time series and subsequently fitting an Interrupted Time Series (ITS) model. 
To quantify toxicity, we compute a score between 0 and 1 for each post using Detoxify ~\citep{hanu_detoxify_2020}, where higher values signify more offensive or hostile language.
Because daily post volume varies widely, we employ a robust bootstrap procedure to obtain stable daily estimates.

For each day, we first draw 100 bootstrap samples with replacement, each consisting of up to 100 posts from that day (sampling with replacement from all available posts if fewer than 100 exist).
We then compute the median toxicity within each of the 100 bootstrap samples before averaging these 100 medians to establish a single, stable toxicity value for that day.
This combination of the bootstrap and median significantly reduces sensitivity to outliers and to very low-volume days compared to simple daily means or medians.
Finally, we standardize the resulting daily series with a z-score transformation so that the values represent the number of standard deviations from the overall mean toxicity.

We fit a segmented linear regression model with ordinary least squares (OLS), treating August 10, 2019 (the day of Epstein's death) as the intervention point $t_0$~\citep{trujillo_make_2022}.
The model is:
\begin{equation}
    y_t = \beta_0 + \beta_1 t + \beta_2 D_t + \beta_3 P_t,
\end{equation}
where $y_t$ is standardized toxicity on day $t$, $D_t$ equals 0 before $t_0$ and 1 afterward (immediate level change), and $P_t$ counts days since $t_0$ (post-event trend). 
The intercept $\beta_0$ is the baseline level, $\beta_1$ is the pre-event trend, $\beta_2$ captures the immediate change at Epstein's death, and $\beta_3$ measures any change in slope after the event.

\subsection{Lexical categories}

We use Empath~\citep{fast_ejhfastempath-client_2025} to measure thematic content in posts. 
We focus on 17 categories that are relevant to conspiracy discourse and community dynamics, including \textit{government, politics, crime, violence, trust, conflict, secrecy,} and \textit{dispute}, as well as a custom \textit{paranoia} category that we build from seed words such as \textit{suspicion, paranoid, watching, surveillance, threat,} and \textit{conspiracy}. 
Empath assigns each post a length-normalized score per category based on the number of category words it contains.

Empath's lexicon derives from general-purpose English and does not fully capture all slang, sarcasm, or coded language common on Reddit.
We therefore treat category scores as approximate indicators of themes rather than exhaustive semantic labels.
We correct for multiple tests via Benjamini-Hochberg false discovery rate control.

We examine lexical shifts in two ways. 
First, we quantify an average \emph{within-user behavioral shift} for core members~\citep{balsamo2023pursuit}.
For each event (Epstein's arrest and death), we define a one-month window before and a one-month window after the event. 
For every core user who posts in both windows, we compute Empath scores in the pre- and post-event months and take the difference. 
We then average these user-level differences by category to obtain the mean within-user shift around each event.

Second, we measure a \emph{newcomer behavioral shift} relative to core members.
For each event, we compute a baseline of core users' Empath scores in the month before the event. 
We then compute mean scores for newcomers in the month after the event and define the shift for each category as the difference between the newcomer mean and the core baseline. 
This shows whether newcomers emphasize different themes than established members when they enter the community.

\subsection{User retention and survival analysis}
We study user retention to understand how long different cohorts remain active in \texttt{r/conspiracy}. 
For each newcomer, we compute the number of days between their first and last observed post in the subreddit. 
We then define an event indicator that marks apparent churn: 
the indicator equals 1 if the user has no activity for at least 30 consecutive days before the end of 2019, and 0 otherwise. 
Users with a churn indicator of 0 are right-censored, because we only know they are active up to the end of the observation window.

As robustness checks on the inactivity threshold, we repeat the analyses with 15- and 60-day inactivity windows.
The 15-day cutoff yields essentially identical results, while the 60-day cutoff causes the two post-arrest cohorts to overlap but preserves the separation between post-death and post-arrest cohorts.

Using these durations and event indicators, we estimate retention with Kaplan-Meier survival curves, one per cohort. 
The estimator computes, at each distinct departure time, the conditional probability that users remain active given that they have not yet left, and multiplies these probabilities over time. 
The resulting curve shows, for each cohort, the probability that a user is still active in \texttt{r/conspiracy} after a given number of days since their first post. 
We also compute the Integrated Brier Score for each cohort to summarize how well the survival model predicts observed departures over time. 
Lower values of IBS indicate that the timing of user churn within a cohort is more regular and closely aligns with the model's survival estimates, whereas higher scores reflect more diverse participation patterns.

\subsection{Linguistic integration}
To analyze linguistic integration, we measure how closely newcomers' posts align semantically with the language of established core users over time~\cite{danescu-niculescu-mizil_no_2013}. 
This analysis uses sentence embeddings from the all-MiniLM-L6-v2 Sentence-BERT model in the SentenceTransformers library.\footnote{\url{https://sbert.net}}
The model maps each post to a 384-dimensional vector by applying the default tokenizer, truncating sequences longer than 512 tokens, and using mean pooling over token representations. 
We use the pre-trained model without fine-tuning, which keeps the analysis reproducible and avoids overfitting to this specific community.

For each week $w$, we compute a centroid embedding by averaging SBERT vectors for all core-user posts in that week
\begin{equation}
    LM_w = \frac{1}{N_w} \sum_{i=1}^{N_w} SBERT(p_i),
\end{equation}
where $N_w$ is the number of core posts in week $w$ and $SBERT(p_i)$ is the embedding of post $p_i$. 
This centroid acts as a reference point for the community's typical language in that week.

For every newcomer post $p$ with embedding $v = SBERT(p)$, we identify the week $w$ in which it appears and compute its cosine distance from the corresponding centroid $u = LM_w$
\begin{equation}
    \mathop{cos\_dist}(u, v) = 1 - \frac{u \cdot v}{\|u\|\,\|v\|}.
\end{equation}
Higher cosine distance indicates the newcomer's language is semantically further from core discourse in that week, while lower distance indicates stronger alignment.
We compute this distance for all newcomer posts and, for each cohort and week, take the mean distance across posts. 
This procedure results in a weekly time series of semantic distance for each cohort. 
We then smooth each series with a three-week rolling mean to filter out noise and highlight longer-term trends.

As a robustness check, we also train weekly bigram language models on core-user posts and compute cross-entropy for newcomer posts under the corresponding weekly model. 
These models yield trajectories that match the cosine-distance trends for three of the four cohorts: 
only the `arrest--not-engaged' cohort shows instability across the two methods. 
We therefore interpret that cohort's semantic trajectory with caution and focus on patterns that are consistent across both methodologies.

\section{Results}
\subsection{Behavioral and thematic responses (RQ1)}

\begin{figure*}[!t]
  \centering
  \includegraphics[width=\linewidth]{images/ITS-toxicity-core.pdf}
  \caption{Interrupted time series analysis of core users' toxicity. 
  The blue line shows the observed daily toxicity estimate. 
  The red segments show the fitted pre-event and post-event trends from the OLS regression. 
  The vertical line at $t_0$ marks Epstein's death. 
  One extreme spike above 7 standard deviations in June comes from a day with very few posts and one highly toxic message; this outlier does not drive the fitted trend because the bootstrap uses medians.}
  \label{fig:its-toxicity}
\end{figure*}

Our first research question asks how the visibility event alters the behavior of established community members. 
The ITS analysis in \Cref{fig:its-toxicity} shows a clear and immediate shift in how core users write. 
Toxicity scores drop by 1.14 standard deviations ($p < 0.001$) on the day of Epstein's death. 
This level change is large and negative, while the post-event slope coefficient is small and not statistically different from zero (\Cref{table:its-toxicity}). 
In the months after the event, daily toxicity values fluctuate around a lower mean than in the pre-event period, without a strong upward rebound in the fitted trend.

\begin{table}[b]
    \label{table:its-toxicity}
\begin{center}
    \sisetup{table-number-alignment=center,table-format=1.3,round-mode=places,round-precision=3}
    \begin{tabular}{lSS}
        \toprule
        \textbf{Term} & \textbf{Coefficient} & \textbf{p-value} \\
        \midrule
        $\beta_0 \; (const)$ & 0.0717  & 0.567 \\
        $\beta_1 \; (t)$     & 0.0019  & 0.053 \\
        $\beta_2 \; (D_t)$   & -1.1377 & 0.000 \\
        $\beta_3 \; (P_t)$   & 0.0010  & 0.624 \\
        \bottomrule
    \end{tabular}
\end{center}
\caption{Interrupted time series regression estimates for core users' toxicity.}
\end{table}

\begin{figure}[!t]
  \centering  \includegraphics[width=\linewidth]{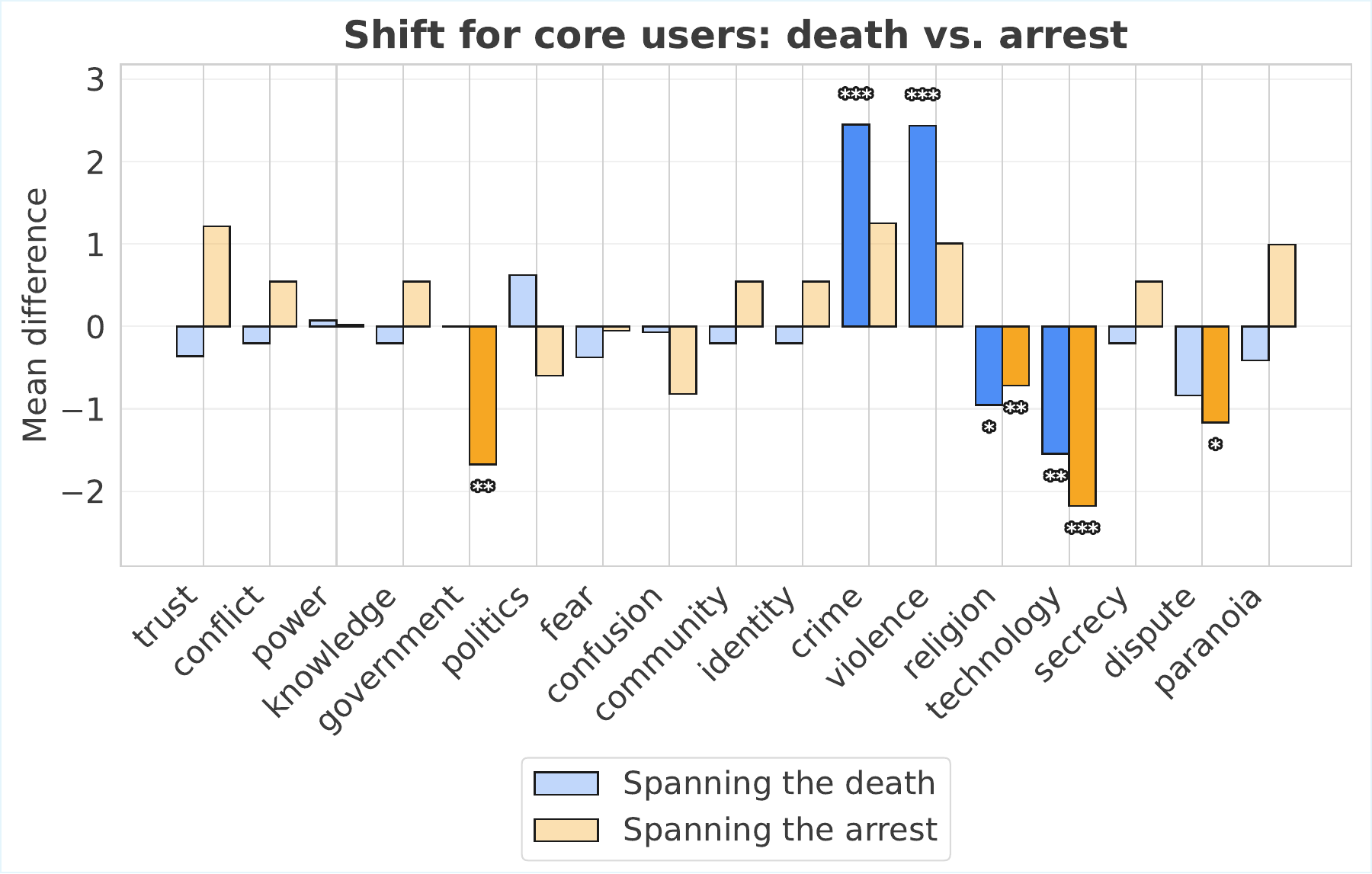}
  \caption{Shift in Empath category of core users' posts spanning the arrest and the death of Jeffrey Epstein in \texttt{r/conspiracy}.  
  Asterisks denote significance levels (***: $p < 0.001$; **: $p < 0.01$; *: $p < 0.05$).}
  \label{fig:empath-shift-core}
\end{figure}

\begin{figure}[!t]
  \centering    \includegraphics[width=\linewidth]{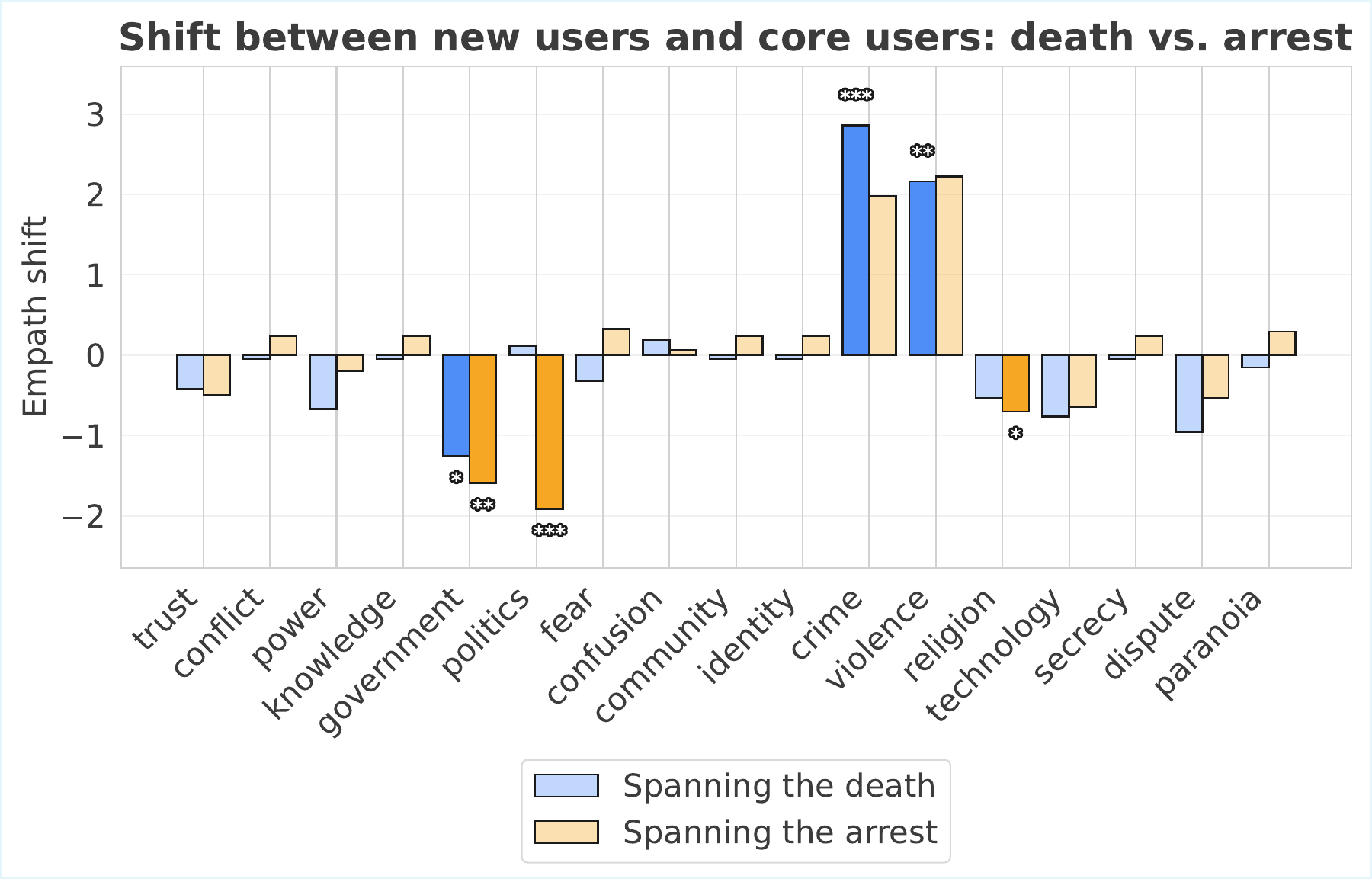}
  \caption{Shift in Empath category between core and new users' posts spanning the arrest and the death of Jeffrey Epstein in \texttt{r/conspiracy}. 
  Asterisks denote significance levels (***: $p < 0.001$; **: $p < 0.01$; *: $p < 0.05$).}
    \label{fig:empath-shift-new-core}
\end{figure}

The lexical analysis with Empath reveals distinct thematic shifts both within the posts of core users over time and between core users and newcomers, as shown in \Cref{fig:empath-shift-core} and \Cref{fig:empath-shift-new-core}.
After Epstein's death, core users increase their use of vocabulary related to crime and violence, while after the arrest, they reduce references to government, dispute, religion, and technology. 
These differences appear as statistically significant within-user shifts in the corresponding categories. 
The patterns show that established members change the topics and frames they emphasize around the two events, even when their overall posting volume remains high.

\Cref{fig:empath-shift-new-core} shows that, after Epstein's death, new users use significantly more crime and violence terms and significantly fewer government terms than core users. After the arrest, the only significant shifts are negative: new users refer to government, politics, and religion less often than core users. 

Overall, the visibility event coincides with a sharp and lasting drop in how toxic core users' posts are.
This result might be due to several factors such as the adaptation on the part of the core users to the influx of newcomers,\footnote{\url{https://reddit.com/r/conspiracy/comments/cpb1ay}} whose toxicity level is reasonably lower, or due to the awareness of increased visibility due to having landed on Reddit's homepage.
At the same time, established members shift their language toward crime and violence and away from government- and religion-related themes, seemingly focusing more on the event at hand.
Newcomers show similar topic choices but lean even more toward crime and less toward institutional politics than core users, which can be interpreted as reticence to engage with speculations and deeper conspiracy theories.
These results show that the event changes both the tone and the main themes in the community's discussions.

\subsection{Community composition and user retention (RQ2)}

The second research question examines how homepage visibility affects users who join the community and how long they stay. 
A Kaplan-Meier survival analysis~\cite{goel2010understanding} reveals that the context of discovery is a strong predictor of retention. 
Users who join during the arrest period, which we use as a proxy for different a more organic discovery context, show higher survival probabilities at almost every point in time than users who join after the death event, corresponding to visibility-driven discovery.

\begin{figure}[!t]
  \centering
  \includegraphics[width=\linewidth]{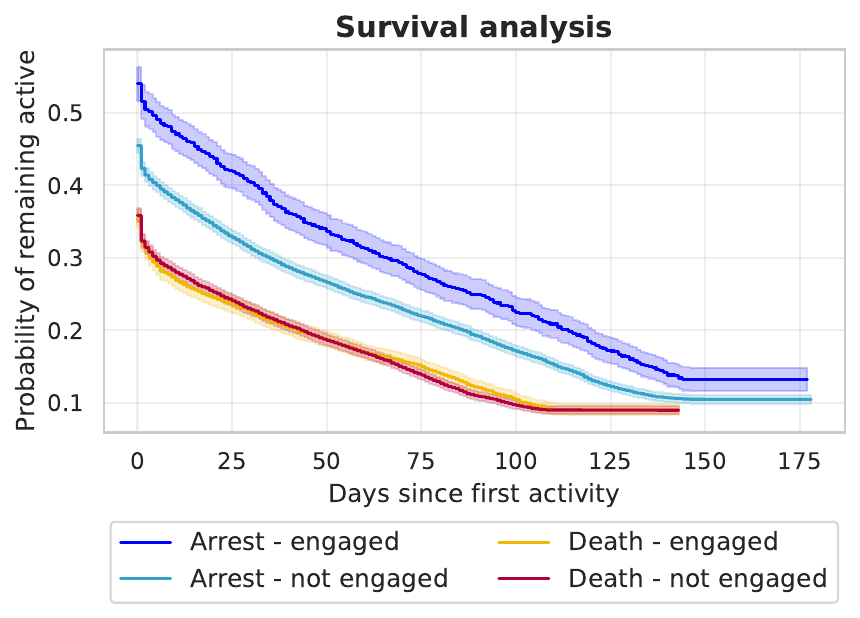}  \caption{Kaplan-Meier survival curves for the four newcomer cohorts. 
  From top to bottom over most of the time range, the curves are: arrest--engaged, arrest--not-engaged, death--engaged, and death--not-engaged. 
  Solid and dashed line styles distinguish cohorts in addition to colour for accessibility.}
  \label{fig:user_retention}
\end{figure}

\Cref{fig:user_retention} shows that, among the arrest-era cohorts, users whose first post is about Epstein have the highest retention. 
Arrest-era users who start in non-Epstein threads leave more quickly but still outlast both death-era cohorts. 
Users who arrive during the amplified mainstream visibility after the death event churn faster, regardless of whether their first post is about Epstein or about other conspiracy topics. 
The two death-era curves almost overlap, which indicates that discovery context matters more for retention than the specific initial topic.

\begin{table}[t]
    
    \centering
    \label{table:ibs-cohorts}
    \begin{tabular}{r@{}lS[table-number-alignment=right,table-format=1.3,round-mode=places,round-precision=3]}
        \toprule
        \multicolumn{2}{l}{\textbf{Cohort}} & \textbf{Integrated Brier Score} \\
        \midrule
        Arrest&--engaged      & 0.18252 \\
        Arrest&--not-engaged  & 0.15542 \\
        Death&--engaged       & 0.13160 \\
        Death&--not-engaged   & 0.13101 \\
        \bottomrule
    \end{tabular}
    \caption{Integrated Brier Scores by cohort}
\end{table}

The Integrated Brier Score (IBS) summarizes how well the survival models track observed departures over time.
Lower values indicate more homogeneous and predictable departure patterns. 
The two post-death cohorts have lower IBS (around 0.131) than the arrest-era cohorts (0.155--0.183).
This result indicates that although death-era users leave sooner, their retention pattern is more regular. 
The higher IBS for arrest-era users reflects more heterogeneous engagement, consistent with a mix of very long-term and short-term participants and a more organic churn process.

These findings reveal that users who discover the conspiracy community after Epstein's arrest stay active much longer than those who arrive after his death, while post-death users churn quickly.
This suggests that the context of discovery shapes not only how long people stay but also how diverse their engagement trajectories are.
Users whose activity coincides with sudden mainstream visibility are likely pushed by the exogenous stimulus of a collective attention spike;
when the stimulus wanes, there is little endogenous engagement left with the community, so the users abandon it in a predictable way.

\subsection{Linguistic integration patterns (RQ3)}

\begin{figure}[t]
    \centering
    \includegraphics[width=\linewidth]{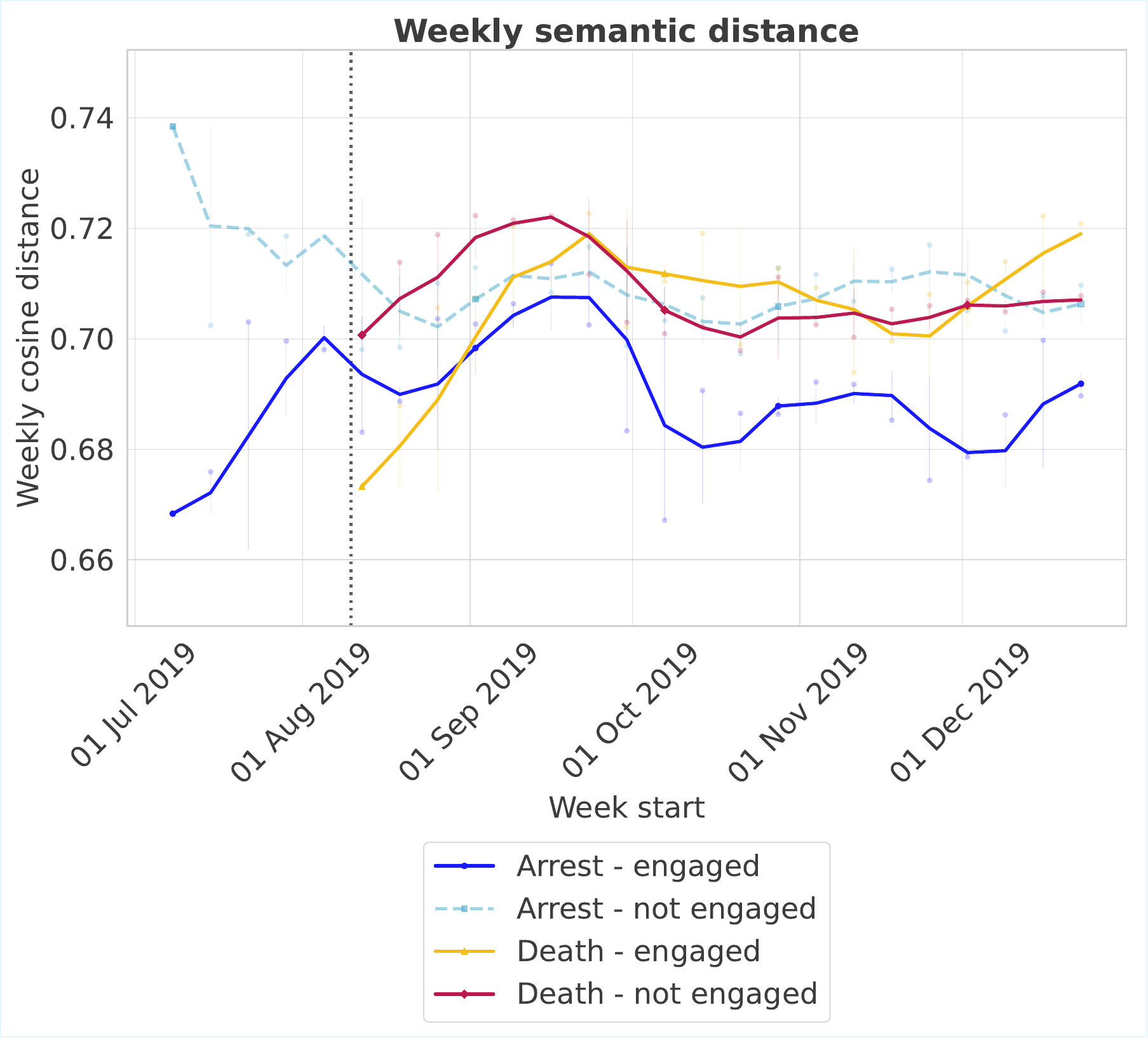}
    \caption{Temporal evolution of semantic alignment for new user cohorts, measured as cosine distance from weekly core-user language centroids. 
    Solid curves show three-week rolling means of weekly average distances; dots show unsmoothed weekly means. 
    The light blue dashed curve indicates the arrest--not-engaged cohort, whose trajectory is sensitive to the choice of metric. 
    The vertical dotted line marks Epstein's death.}
    \label{fig:sbert}
\end{figure}

The third research question asks whether the timing and context of entry influence how well new users integrate linguistically. 
The semantic distance analysis uses Sentence-BERT embeddings to track how close newcomer posts are to the weekly language of core users. 
\Cref{fig:sbert} plots the weekly mean cosine distance for each cohort; 
values fall in a narrow band between roughly $0.66$ and $0.74$, so absolute effect sizes are modest, and the focus is on the relative ordering of cohorts over time.

Users who arrive during the arrest period and discuss Epstein have the lowest cosine distance to core discourse and maintain this characteristic throughout the observation window. 
Their weekly trajectories show an initial convergence followed by stable and consistent lower semantic distance over time. 
Death-era cohorts, regardless of topic, stay at higher distances; 
they do not close the gap with the arrest--engaged cohort over the five months. 
The arrest--not-engaged cohort remains the farthest from the core and is the only cohort whose trajectory changes noticeably when using cross-entropy from bigram language models instead of cosine distance with SBERT (which is why it appears as a dashed line in \Cref{fig:sbert}).

Overall, the cosine distances vary within a narrow range, so the differences between cohorts are modest in absolute terms.
Even so, users who arrive during the arrest period and post about Epstein use language that is consistently closest to that of core members.
Death-era and non-engaged arrest-era users stay farther from core discourse over time, which suggests a possible weaker linguistic integration of this cohort.
We observe that users with higher semantic distance also tend to have shorter retention, although causality cannot be established.
\section{Discussion}
To understand how a sudden visibility shock like Epstein's death reshaped participation and discourse in a conspiracy community, we have analyzed all posts and comments in \texttt{r/conspiracy} throughout 2019 and constructed four newcomer cohorts based on when and where users first engaged with Epstein-related content. 

We have measured toxicity over time and applied an Interrupted Time Series model to quantify changes among core users around Epstein's death; this analysis revealed a sharp and sustained drop in toxicity immediately following the visibility spike. 
We have examined thematic dynamics using Empath lexical categories and found shifts toward crime- and violence-related language, alongside declines in government and religion framing. 
We have modeled newcomer retention through Kaplan-Meier survival curves and observed that users who activate during homepage exposure churn faster and more uniformly than those joining organically during the arrest period. 
Finally, we have evaluated linguistic integration using Sentence-BERT embeddings and bigram cross-entropy, showing that organic arrest-period newcomers converge most closely with core discourse, while post-death cohorts remain semantically distant throughout the study window.

Our findings reveal several insights into how the mainstream visibility event of being featured on Reddit's homepage reshaped \texttt{r/conspiracy}.
First, the toxicity analysis shows that core users reduced offensive language immediately after the spike and maintained lower toxicity levels without a clear rebound. 
This suggests that established members could have adjusted their style when the community becomes more publicly exposed---whether through self-moderation, moderation enforcement, language mirroring, or social pressure from a broader audience.

Second, the retention and IBS results indicate that discovery context functions as a compositional filter. Homepage visibility is associated with large influxes of short-lived participants who behave in a relatively uniform way, whereas organic discovery during the arrest period brought in fewer but more varied users, including those who remain active long-term. Combined with semantic distance findings, this result suggests that the visibility spike expands the audience but not the pool of deeply integrated community members.

Third, the semantic integration analysis shows stable and persistent differences between newcomers.
Arrest--engaged users remain most aligned with core discourse, while death-era cohorts stay consistently farther away.
Cross-entropy trajectories reinforce this pattern for three of the four cohorts.
In this context, mainstream visibility has an impact on who arrives and how they write, but it does not fundamentally shift the linguistic center of the community.

These patterns contribute to debates on mainstream exposure, conspiracy consumption, and radicalization---within the limits of observable posting activity.
During this five-month window, most users who joined during homepage visibility do not remain active or converge toward core linguistic norms.
We do not observe evidence that incidental exposure alone is associated with large numbers of durable, linguistically integrated participants.
Rather, users who discover the conspiracy community during the arrest era, and who might have an already-existing intrinsic motivation, are the ones that persist the longest~\citep{corso2025early}.
This result does not preclude the existence of other forms of radicalization that occur privately, later, or off-platform; it simply indicates that such processes are not reflected in posting behavior in this dataset.

\subsection{Limitations and future work}
Like any empirical work, this study is not exempt from limitations.
The primary limitation is the observation window. 
The analysis covers only five months around the events related to Epstein and therefore focuses on a period of acute crisis rather than on longer arcs of community evolution. 
Integration and retention patterns may look different when the community is not under intense public attention, when other news dominates the public agenda, or when the subreddit experiences slower, more endogenous change.
However, this event gives us an ideal case study experiment for the effects of mainstream visibility that is invaluable to study the dynamics of opinion formation on social media.

A second limitation is that the design does not rule out alternative explanations for the observed gaps between cohorts. 
Other events in 2019, such as concurrent political developments or media cycles, may influence who visits \texttt{r/conspiracy} and how they behave. 
Subreddit-specific moderation decisions around the time of the Epstein spike, such as stricter enforcement of rules or the removal of especially toxic threads, could also contribute to shifts in toxicity and engagement that we attribute to visibility. 
Because the analysis uses only public post data, it cannot directly measure these background processes.

Discovery pathways are also inferred rather than directly observed. 
We approximate homepage exposure with temporal and topical cohorts together with observational data drawn from Google Trends, assuming that the sharp spike in activity after August 10 reflects the subreddit's placement on Reddit's homepage. 
In reality, some users in the `death' cohorts may have arrived via organic browsing or external links, and some `arrest' users may have seen recommendations or cross-posts. 
This potential misclassification likely blurs the differences between cohorts rather than producing them, but it still weakens the causal force of the conclusions.

The availability and choice of specific measurement tools introduce further constraints. 
Toxicity and lexical categories come from models trained on general English and may miss context-specific slang, sarcasm, or coded language common in conspiracy communities.
Sentence embeddings rely on a pre-trained SBERT model that may underweight very short posts, strip URLs and usernames into generic tokens, and compress different senses of the same word into a single vector. 
As a result, the semantic distance measures capture broad alignment rather than fine-grained ideological nuance.
Using a mix of fine-tuned embeddings, large language models, and human validation could be a viable strategy to improve the reliability and accuracy of the current results.
Moreover, we cannot distinguish whether these differences reflect post-entry linguistic adaptation or pre-existing similarity between newcomers and core users (selection effects).

Future work can address these gaps in several ways. 
Extending the tracking period to at least a year would reveal whether late arrivals eventually integrate or remain distinct once the immediate attention spike fades. 
Incorporating external data on news cycles or platform-level changes could help separate the effect of Epstein-specific visibility from broader political or media trends. 
Comparing similar visibility events across other subreddits and platforms would test how general the compositional filter effect is. 
Finally, combining behavioral and linguistic signals with network analysis and, where possible, richer platform metadata would clarify whether visibility-driven users form separate clusters, attach to core members, or simply drift away without building ties.

\subsection{Ethical use of data and consent}
This study analyzes discussion data from the subreddit \texttt{r/conspiracy} that was collected via the public Pushshift Reddit archives. 
The dataset consists solely of posts and comments that were already publicly visible on Reddit at the time of collection, and no direct contact with Reddit users took place. 
To protect user privacy, we conduct all analyses on de-identified records and report only aggregate results.
No usernames, direct identifiers, or verbatim quotations that could enable re-identification are included in this paper. 
Rediscovery or amplification of harmful conspiratorial content is a potential risk in this domain; accordingly, we avoid reproducing sensitive text, refrain from linking specific posts or users, and restrict analysis to behavioral and linguistic aggregates. 
Data handling followed institutional ethical standards for research on public online communities.
We leave the repository with the data and the code to replicate our analyses as follows: \url{https://github.com/aattanasio/Paper.git}
\bibliography{references}

\section*{Paper Checklist}

\begin{enumerate}

\item For most authors...
\begin{enumerate}
    \item  Would answering this research question advance science without violating social contracts, such as violating privacy norms, perpetuating unfair profiling, exacerbating the socio-economic divide, or implying disrespect to societies or cultures?
    \answerYes{Yes.}
  \item Do your main claims in the abstract and introduction accurately reflect the paper's contributions and scope?
    \answerYes{Yes.}
   \item Do you clarify how the proposed methodological approach is appropriate for the claims made? 
    \answerYes{Yes, in ``Methods''.}
   \item Do you clarify what are possible artifacts in the data used, given population-specific distributions?
    \answerYes{Yes, we discuss biases and limitations in ``Limitations and Future work''}
  \item Did you describe the limitations of your work?
    \answerYes{Yes, limitation are presented and discussed in ``Limitations and Future work''}
  \item Did you discuss any potential negative societal impacts of your work?
    \answerYes{We discuss negative societal impact in ``Ethical Statement''.}
      \item Did you discuss any potential misuse of your work?
    \answerYes{Yes, we discuss potential misuse in ``Ethical use of data and consent''.}
    \item Did you describe steps taken to prevent or mitigate potential negative outcomes of the research, such as data and model documentation, data anonymization, responsible release, access control, and the reproducibility of findings?
    \answerYes{Yes, we describe the details in ``Ethical use of data and consent''.}
  \item Have you read the ethics review guidelines and ensured that your paper conforms to them?
    \answerYes{Yes.}
\end{enumerate}

\item Additionally, if your study involves hypotheses testing...
\begin{enumerate}
  \item Did you clearly state the assumptions underlying all theoretical results?
    \answerNA{NA}
  \item Have you provided justifications for all theoretical results?
    \answerNA{NA}
  \item Did you discuss competing hypotheses or theories that might challenge or complement your theoretical results?
    \answerNA{NA.}
  \item Have you considered alternative mechanisms or explanations that might account for the same outcomes observed in your study?
    \answerNA{NA}
  \item Did you address potential biases or limitations in your theoretical framework?
    \answerNA{NA}
  \item Have you related your theoretical results to the existing literature in social science?
    \answerNA{NA}
  \item Did you discuss the implications of your theoretical results for policy, practice, or further research in the social science domain?
    \answerNA{NA}
\end{enumerate}

\item Additionally, if you are including theoretical proofs...
\begin{enumerate}
  \item Did you state the full set of assumptions of all theoretical results?
    \answerNA{NA.}
	\item Did you include complete proofs of all theoretical results?
    \answerNA{NA.}
\end{enumerate}

\item Additionally, if you ran machine learning experiments...
\begin{enumerate}
  \item Did you include the code, data, and instructions needed to reproduce the main experimental results (either in the supplemental material or as a URL)?
    \answerYes{Data is reported in ``Methods", code will be shared upon acceptance.}
  \item Did you specify all the training details (e.g., data splits, hyperparameters, how they were chosen)?
    \answerNA{NA}
     \item Did you report error bars (e.g., with respect to the random seed after running experiments multiple times)?
    \answerNA{NA}
	\item Did you include the total amount of compute and the type of resources used (e.g., type of GPUs, internal cluster, or cloud provider)?
    \answerNA{NA}
     \item Do you justify how the proposed evaluation is sufficient and appropriate to the claims made? 
    \answerNA{NA}
     \item Do you discuss what is ``the cost`` of misclassification and fault (in)tolerance?
    \answerNA{NA.}
  
\end{enumerate}

\item Additionally, if you are using existing assets (e.g., code, data, models) or curating/releasing new assets, \textbf{without compromising anonymity}...
\begin{enumerate}
  \item If your work uses existing assets, did you cite the creators?
    \answerYes{Yes, we provide citations for all external assets we used.}
  \item Did you mention the license of the assets?
    \answerNA{NA}
  \item Did you include any new assets in the supplemental material or as a URL?
    \answerNA{NA}
  \item Did you discuss whether and how consent was obtained from people whose data you're using/curating?
    \answerNA{NA.}
  \item Did you discuss whether the data you are using/curating contains personally identifiable information or offensive content?
\answerYes{Yes, we describe the details in ``Ethical use of data and consent''}
\item If you are curating or releasing new datasets, did you discuss how you intend to make your datasets FAIR?
\answerNA{NA.}
\item If you are curating or releasing new datasets, did you create a Datasheet for the Dataset? 
\answerNA{NA.}
\end{enumerate}

\item Additionally, if you used crowdsourcing or conducted research with human subjects, \textbf{without compromising anonymity}...
\begin{enumerate}
  \item Did you include the full text of instructions given to participants and screenshots?
    \answerNA{NA.}
  \item Did you describe any potential participant risks, with mentions of Institutional Review Board (IRB) approvals?
    \answerNA{NA.}
  \item Did you include the estimated hourly wage paid to participants and the total amount spent on participant compensation?
    \answerNA{NA.}
   \item Did you discuss how data is stored, shared, and de-identified?
   \answerNA{NA.}
\end{enumerate}

\end{enumerate}

\appendix
\section{Appendix}
As a robustness check for the SBERT-based semantic distance analysis, we also measure alignment using weekly bigram language models and cross-entropy. 
For each week, we train a snapshot language model on all tokenized posts authored by core users, using bigrams with Katz back-off smoothing to estimate probabilities for seen and unseen sequences. 
Given a newcomer post \(p\) consisting of bigrams \(b_1, \ldots, b_N\) and the snapshot model for its week \(w_p\), we compute cross-entropy as
\begin{equation}
    H_p = - \frac{1}{N} \sum_{i=1}^{N} \log P_{SLM_{w_p}}(b_i),
\end{equation}
where \(P_{SLM_{w_p}}(b_i)\) is the bigram probability under the corresponding weekly model. 
Lower cross-entropy indicates that the post uses bigrams that are more typical of core-user language in that week, while higher values reflect linguistic divergence. 
As in the SBERT analysis, we aggregate individual post scores into weekly cohort means and apply a three-week rolling mean to obtain smooth trajectories, which we use only to validate that the ordering of cohorts over time does not depend on the choice of semantic metric. 
The only trajectory that meaningfully differs between cosine-distance and cross-entropy is the cohort of users who join after the arrest but whose first post is not about Epstein, while the other three cohorts show consistent patterns across both methods.

\subsection{Additional Figures}
\begin{figure}[h]
  \centering
  \includegraphics[width=.8\linewidth]{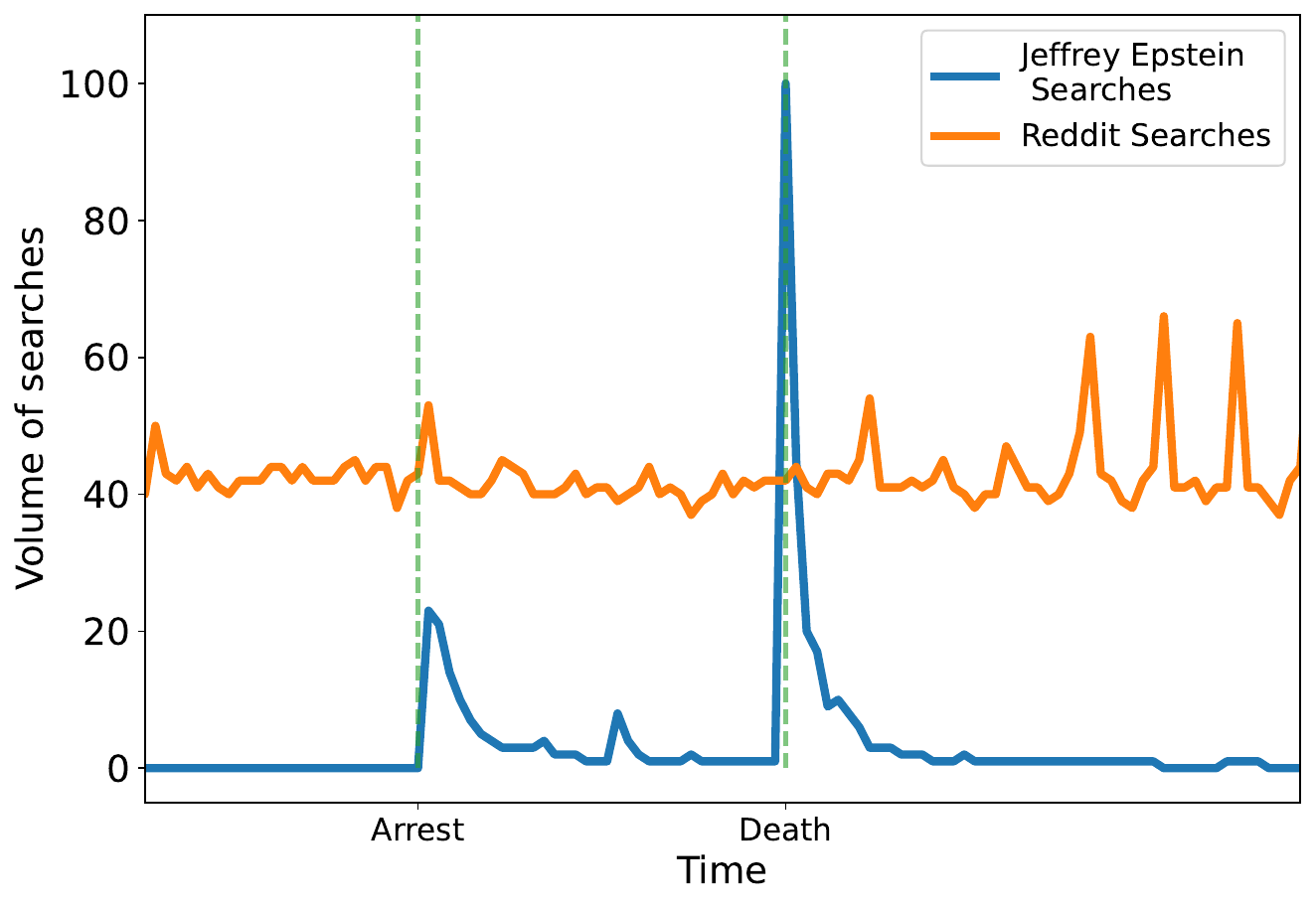}
  \caption{Longitudinal Google Search prevalence of keywords: `Jeffrey Epstein' and `Reddit' during the arrest and death periods.}
  \label{fig:google_searches}
\end{figure}

\begin{figure*}[!t]
  \centering
  \includegraphics[width=1\linewidth]{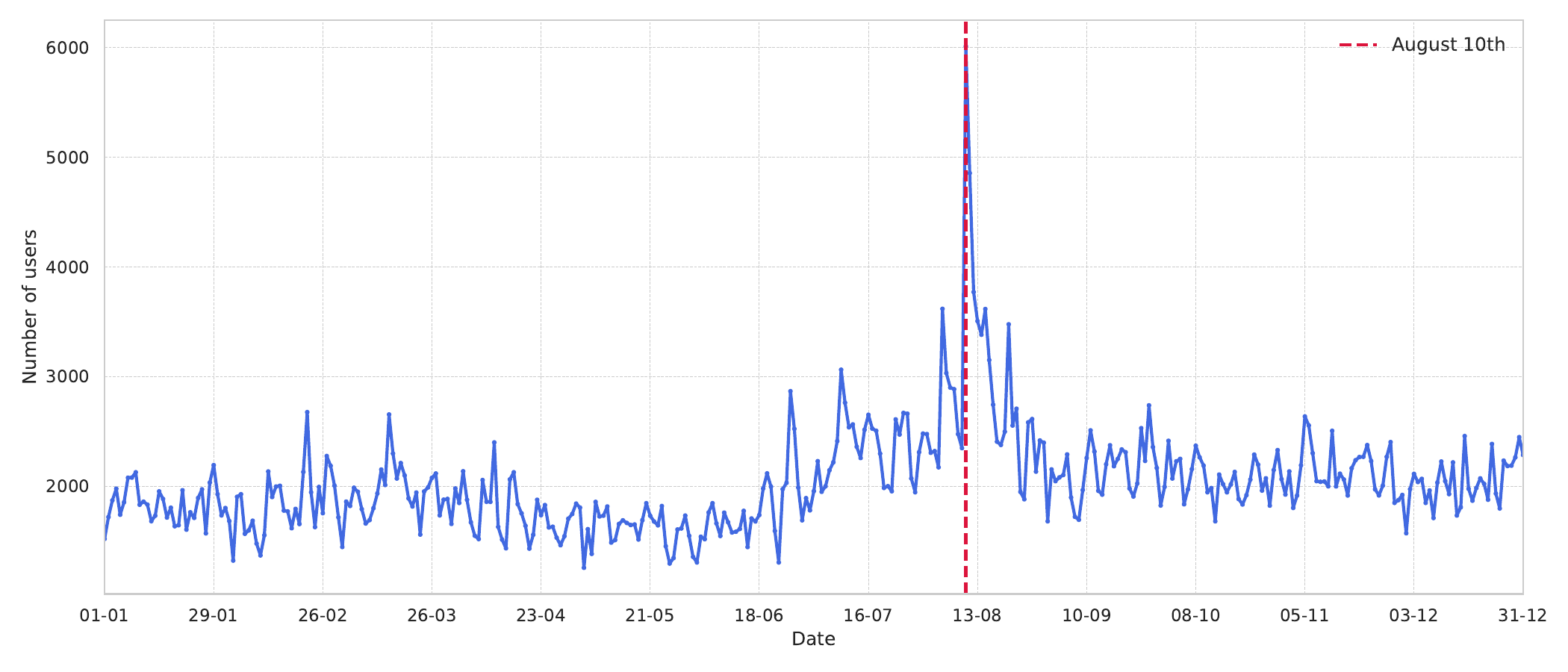}
  \caption{Unique users active on \texttt{r/conspiracy} during the period of our study.}
  \label{fig:app_users_active}
\end{figure*}

In \Cref{fig:google_searches} and \Cref{fig:app_users_active} we show two different measurement of the exogenous event that caused the shock in the community: the death of Jeffrey Epstein.
With the former, we compare the prevalence of the keywords 'Jeffrey Epstein' and 'Reddit' on Google Trends during the two periods of interest. 
We see how the death event provoked a much higher volume of researches.
During the arrest era, both Jeffrey Epstein and Reddit see spikes of interest, possibly linked to users actively searching for further information or discussion on this topic on Reddit.
While for the case of the death era, we see a significant spike of interest related to Jeffrey Epstein, but the same cannot be said for Reddit, which search volume remains mostly unchanged during that period.
In \Cref{fig:app_users_active} we show the unique active users during the period of our study.
It's evident how the death event brought a large volume of unique users to become active in the community, which is a different behavior from what can be observed a month prior for the arrest.
\end{document}